\documentclass[10pt,a4paper]{article}
\usepackage{amsmath,amsfonts,amssymb,graphicx}

\begin{document}

\begin{flushright}
{\small LPTh-Ji 10/001}
\end{flushright}

\begin{center}
\vspace{1.8cm}

{\LARGE{\textit{The Restoration of the Electroweak Symmetry at High
Temperature for Little Higgs}}} \vspace{1.3cm}

\textbf{Amine Ahriche}

\vspace{0.6cm}

\textit{Laboratory of Theoretical Physics, Department of Physics, University
of Jijel, PB 98 Ouled Aissa, DZ-18000 Jijel, Algeria.}
\end{center}

\vspace{1.2cm}

\hrule \vspace{0.5cm}{\textbf{{\large {Abstract}}}}

\vspace{0.3cm}

In this letter, we show that the electroweak symmetry is restored at high
temperature for Little Higgs (LH), when including dominant higher order
thermal corrections, that are consequence of the non-linear nature of the
scalar sector. This leads us to suggest that the LH requires a UV completion
above the scale $\Lambda\lesssim f$.\vspace{0.4cm}\hrule\vspace{1cm}

A promising approach had been proposed to solve the hierarchy problem, where
the Higgs field manifests as a pseudo-Goldstone boson since 3 decades ago
\cite{LHm}. This idea was revived by the development of the Little Higgs (LH)
\cite{LH}. The quadratic corrections (QCs) to the Higgs mass are canceled by
the contributions of new introduced fields as in supersymmetry. In these
models bosonic corrections cancel each other; and similarly for fermions. This
is imposed by the presence of such a global symmetry that is broken
spontaneously by a new scalar vev ($f\sim$1- 10 TeV); in a smart way
(collective breaking), where the Higgs boson mass remains protected from
one-loop QCs above a cut-off scale ($\Lambda\sim4\pi f$), that is estimated
using naive dimensional analysis (NDA). The scalar fields correspond to the
broken generators of the global symmetry; and the scalar potential is just a
combination of effective operators whose gauge and Yukawa origins. The
electroweak symmetry breaking (EWSB) is triggered by large one-loop Yukawa
contributions \cite{LH}. The so-called Littlest Higgs \cite{LH}, which is
based on a spontaneously broken global symmetry $SU(5)$ to $SO(5)$, has been
shown to be phenomenologically consistent \cite{phlh}.

However the EW symmetry restoration at temperature seems to be problematic as
shown for the Littlest Higgs \cite{ELR}. This can be understood due to the
fact that thermal corrections ($\sim T^{2}/12$); and quadratic corrections
($\sim3\Lambda^{2}/16\pi^{2}$) are generated at one-loop from the same
interactions in any gauge theory. This means that the Higgs thermal
corrections will cancel each other also; and the electroweak symmetry will not
be restored at high temperature. Indeed, it was shown in \cite{ELR}, that
above such critical temperature $T_{c}\sim f$, the thermal corrections become
negative and the absolute minimum ($\left\langle h\right\rangle \neq0 $), gets
deeper at higher temperatures instead of being relaxed to zero.

In this work, taking the Littlest Higgs as an example, we try to understand
this unusual behavior of the effective scalar potential. We will check whether
this feature is intrinsic for Little Higgs or just a consequence of incomplete
computations, and therefore the symmetry can be restored at high temperatures.

First let us briefly review the Littlest Higgs model. It is based on an
$SU(5)/SO(5)$ nonlinear sigma model, where the $SU(5)$ symmetry is
spontaneously broken down to $SO(5)$ by a vacuum expectation value of a
$5\times5$ symmetric matrix scalar field, $\Sigma_{0}$. The $SU(5)$ subgroup
($SU(2)\times U(1)$)$_{1}\times$($SU(2)\times U(1)$)$_{2}$; is gauged, its
diagonal subgroup being the SM electroweak group $SU(2)_{L}\times U(1)_{Y}$,
and the axial generators correspond to new heavy gauge bosons. The global
$SU(5)$ symmetry breaking results 14 Goldstone bosons: 4 are identified to be
the Standard Model (SM) Higgs doublet, 6 as complex triplet and 4 as the
Goldstone bosons that give masses to the new heavy gauge bosons. These scalar
degrees of freedom are represented in the nonlinear representation as%
\begin{equation}
\Sigma=e^{2i\pi_{k}X_{k}/f}\Sigma_{0},
\end{equation}
where $\pi_{k}$\ are the 14 Goldstone bosons, $X_{k}$\ are the $SU(5)$ broken
generators, and $f$ is the value of the vev that breaks the global symmetry.
The quark sector also involves a new heavy singlet quark $U$, with two
Yukawa\ couplings $\lambda_{1,2}$, which related to standard Yukawa coupling
as $1/\lambda_{t}^{2}=1/\lambda_{1}^{2}+1/\lambda_{2}^{2}$. Similar relations
hold for the gauge couplings $g_{1,2}$\ and $g_{1,2}^{\prime}$: $1/g^{2}%
=1/g_{1}^{2}+1/g_{2}^{2}$ and $1/g^{\prime2}=1/g_{1}^{\prime2}+1/g_{2}%
^{\prime2}$.

The scalar potential is a summation of effective mass operators due to the
gauge and Yukawa interactions, which is given a la Coleman-Weinberg by
\cite{CW}%
\begin{equation}
V_{CW}(\Sigma)=a_{V}\mathrm{Tr}\left[  M_{g}^{2}\right]  +a_{F}\mathrm{Tr}%
\left[  M_{Y}\cdot M_{Y}^{\dag}\right]  ,
\end{equation}
where $a_{V}$, $a_{F}\sim\mathcal{O}\left(  1\right)  $\ are unknown
parameters associated with these effective operators; their values depend on
the UV completion of the theory. The theory ground state is stable in both of
$h$- and $\phi$-directions if $\Theta=a_{V}\left[  g_{1}^{2}+g_{2}^{2}%
+g_{1}^{\prime2}+g_{2}^{\prime2}\right]
/4-a_{F}\lambda_{1}^{2}/2>0$, and the EW symmetry is not broken,
where h and \ are the neutral components of the doublet and triplet,
respectively. But when including the one-loop corrections
(especially the Yukawa's), the EW symmetry gets broken and the SM
fields acquire masses \cite{LH}.

In Ref. \cite{ELR}, it was shown in figures (1) and (2); how the EW symmetry
gets broken in a slight way after including the one-corrections. While in
figure (3), it was shown how thermal corrections help to relax the minimum to
zero especially for low temperature values ($T\leq0.44f$); but the maximum of
the potential (at $h=\pi f/\sqrt{2}$) is getting down when increasing the
temperature until becomes degenerate together with the absolute minimum at a
critical temperature $T_{c}\sim0.96f$. Above this temperature value ($T>f$),
this new minimum gets deeper and deeper when increasing temperature. This
unusual behavior had lead to the conclusion that symmetry can not be restored
at high temperatures in these models.

Before explaining this behavior, let's address the following comment on the
EWSB. The realization of a slight EW symmetry (as in figures (1) and (2) of
\cite{ELR}) is practically impossible, unless taking the cut-off value less
than the NDA value ($\Lambda<4\pi f$).\ This leads to ask questions about the
LH breaking scale $\Lambda$, and\ whether the NDA value is the correct one?
For the cut-off NDA value, the negative Yukawa one-loop corrections at the
absolute minimum, are significantly large with respect to the tree-level
potential (or the parameter $\Theta$), which pushes away the Higgs vev from
its desired value\footnote{In our work, we have chosen the parameters in a way
we can compare our results with those of \cite{ELR}. One should notice that
the authors in \cite{ELR} have used the $SU(2)\times SU(2)\times U(1)$ model
instead the $\left[ SU(2)\times U(1)\right]  ^{2}$ one, but the effect of the
missing $U(1)$ is small and then the comparison is still meaningful.}. But
when reducing its value until $\Lambda\sim1.1f$, the Higgs vev is given
exactly by the SM value, $\upsilon=246$ GeV, for $f=1$ TeV \footnote{If we
take $f=2$ TeV ($f=10$ TeV), the cut-off $\Lambda$\ should be taken
$\sim0.91f$ $\left(  \sim0.86f\right)  $ in order to get $\upsilon=246$ GeV.};
and the Yukawa contribution is comparable to $V_{CW}$. This bound is
consistent with the suggestion of the cut-off value $\Lambda\leq\Lambda
_{NDA}/\sqrt{20}\sim2.8f$; that is coming from the study of the scalar loops
in the Littlest Higgs \cite{esno}. Indeed, that is not the only hint to push
the cut-off scale below the NDA value, but another bound comes from the
unitarity violation where it was suggested that $\Lambda\sim(3-4)f$ \cite{Uni}.

The scalar potential behavior at high temperature that is mentioned in
\cite{ELR}\ can be understood by estimating the field-dependant masses,
especially the Yukawa's, within one period $h\in\lbrack0,\sqrt{2}\pi f]$.
Naively, the gauge field masses vary below the squared values of the gauge
couplings in units of $f^{2}$, and scalar masses vary below a combination of
$a_{V}g_{i}^{2}$ and $a_{F}\lambda_{1}^{2}$ in units of $f^{2}$. These masses
are significantly small when compared with the two Yukawa eigenmasses, the
lightest one lies below $\lambda_{t}^{2}$, and the heaviest one lies between
$\lambda_{2}^{2}$ and $\lambda_{1}^{2}+\lambda_{2}^{2}$ in units of $f^{2}$.
According to these values, the thermal integral \cite{Th}, that involves
$\left(  m^{2}/T^{2}\right)  $ as a variable, will not be suppressed for all
fields; and the unsuppressed fermionic contributions, that have a large
negative multiplicity ($n_{t}=n_{T}=-12$), will dominate the thermal scalar
potential since they are $T^{4}$-proportional. In other gauge theories, SM as
an e.g.,\ most of the fields contributions are suppressed for large values of
the scalar field $h$, but this not the case here because the scalar sector is periodic.

However, one need to comment on the problem of the vacuum misalignment induced
by top quark in LH models \cite{TMa}. In order that the EW vacuum would be the
deepest one, some constrains on the gauge couplings are imposed, these
constrains become less serious as the cut-off is taken to be less then the NDA
value. In the wrong vacuum alignment, that leads to the breaking ($SU(2)\times
U(1)$)$^{2}\rightarrow$($U(1)$)$^{2}$\ instead of ($SU(2)\times U(1)$%
)$^{2}\rightarrow SU(2)\times U(1)$, the Yukawa eigenmasses are of order
$\lambda_{1,2}^{2}$, therefore their thermal contribution to the effective
potential will be always less than in the EW vacuum, this means that the EW
ground state is always the deepest at finite temperature.\footnote{Indeed, a
careful study should be performed to check if there are some parameter regions
where the wrong vacuum is the deepest one at some temperature values,
especially when introducing the corrections (\ref{mto}).}

Then one concludes that this behavior is a consequence of periodicity of the
non-linear nature of the scalar representation; in addition to the largeness
of negative fermionic contributions to the thermal effective potential, that
can not be compensated by other bosonic contributions. But were all the
significant contributions taken into account to obtain this behavior?

\begin{figure}[h]
\begin{center}
\includegraphics[width=9cm,height=5cm]{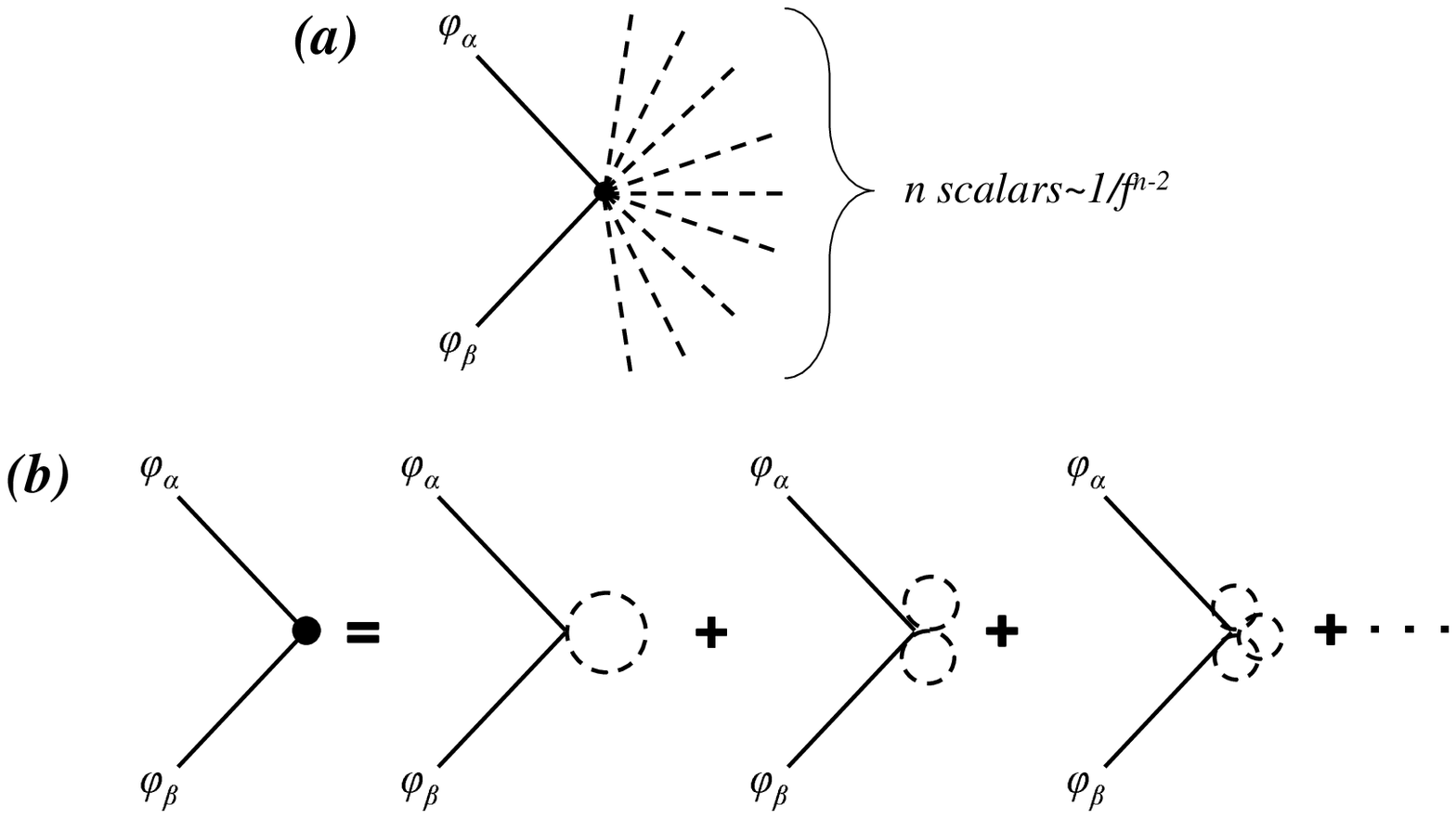}
\end{center}
\caption{\textit{In Fig. a, the fields $\varphi_{\alpha}$ and $\varphi_{\beta
}$ could be scalars, gauge fields or a fermion-antifermion pair. These novel
vertices could result higher order loops corrections to the mass-squared
matrix element $M_{\alpha\beta}^{2}$ as shown in Fig. b.}}%
\label{nl}%
\end{figure}

The nonlinear nature of the scalar sector implies the existence of a new type
of interactions like in Fig. \ref{nl}-a, when expanding the field matrix
$\Sigma$ in powers of $1/f$ in the Lagrangian. These vertices, which are
suppressed as $1/f^{n-2}$, could result higher order loop corrections as shown
in Fig. \ref{nl}-b.

These corrections, in Fig. \ref{nl}-b, are not the only possible contractions
of the vertices in Fig. \ref{nl}-a, but they are the dominant contributions at
high temperature, since each scalar loop gives $T^{2}/12$. Therefore, they
lead to a thermal correction to the field mass-squared matrix element
[$\alpha,\beta$], of the form
\begin{equation}
m^{2}(T)\sim m^{2}+T^{2}\sum\nolimits_{n}c_{n}\left(  T^{2}/f^{2}\right)
^{n},\label{mto}%
\end{equation}
where the zeroth order corresponds to the usual thermal corrections. The
computation of the parameters\ $c_{n}$ for each mass-squared matrix element
tends to determine the vertices of the interactions with scalar degrees of
freedom \cite{next}, where two temperature regimes could be defined: the first
regime where below such temperature $T<f$, only the light degrees of freedom
contribute to the $c_{n}$ parameters, and the second one $T>f$, where all
degrees of freedom are in thermal equilibrium. This type of contributions
could be very important at temperatures $T\gtrsim f$ if the $c_{n}$ parameters
are not largely suppressed. One can take these novel corrections into account
by doing such a resummation with the field-dependant masses in the thermal
integral \cite{Th}, are replaced by thermally corrected masses (\ref{mto}).
This is the same way how the daisy contribution \cite{ring}, was introduced
into the effective potential in order to increase\ the cubic term, and
therefore the region of field values where the Higgs and Goldstone bosons
self-energies suffer from IR divergences, will be decreased. In the high
temperature approximation, this replacement affects mainly the cubic term, and
therefore only bosonic fields will be taken into account. Since we have light
and heavy particles in our model, and we want to take all of them into account
by using the exact formula of the thermal integral \cite{Th}, we will replace
also the fermionic thermally corrected masses to evaluate the effective
thermal correction.

\begin{figure}[h]
\begin{center}
\includegraphics[width=8cm,height=6cm]{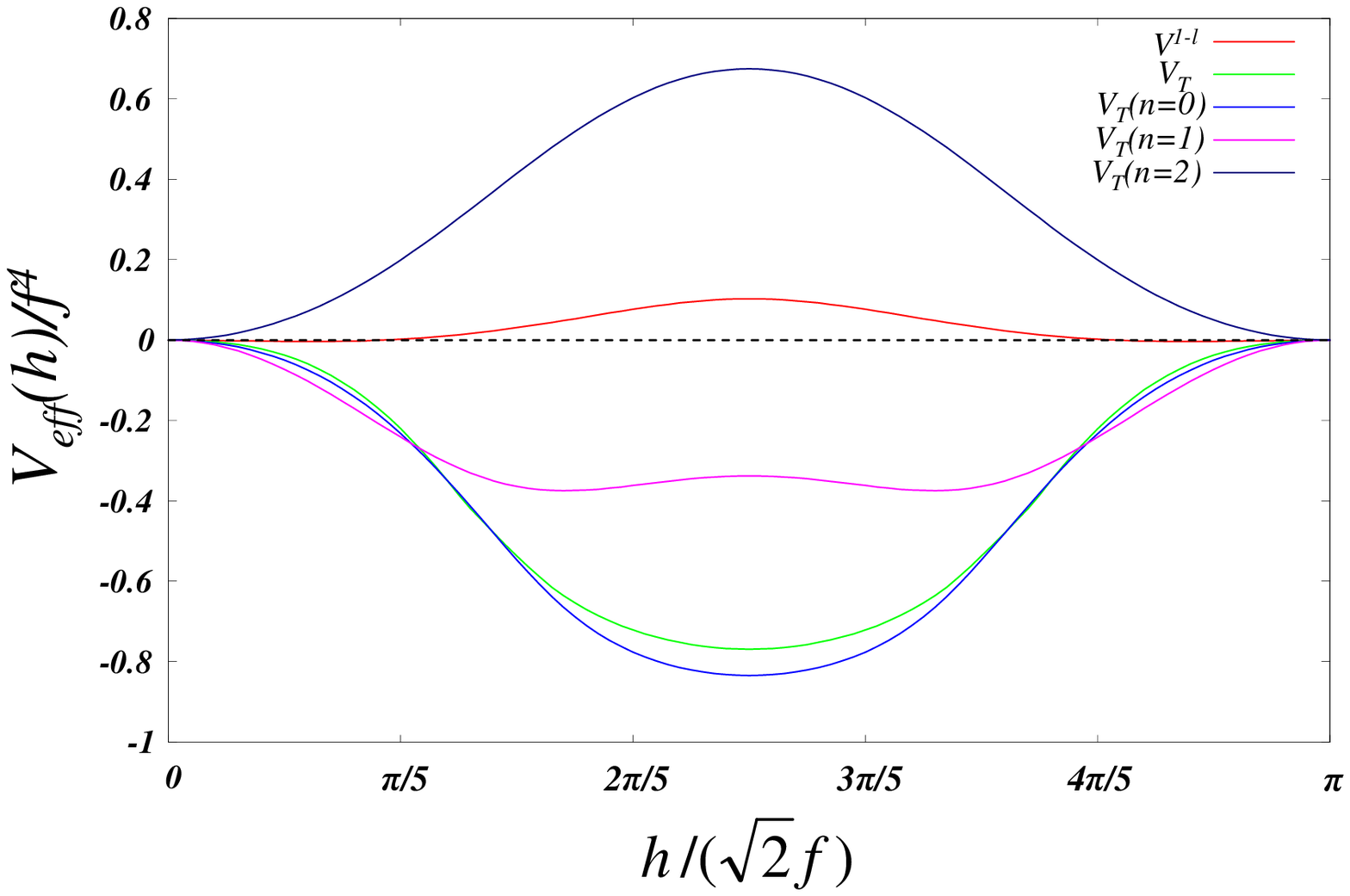}
\end{center}
\caption{\textit{The effective potential at T=0, and at T=1,7f computed in the
standard way, and using the resumed thermal masses (\ref{mto}) taking into
account 1-loop (n=0), 2-loop (n=1), and 3-loop (n=2) corrections.}}%
\label{Vt}%
\end{figure}

Now, the question is how these corrections can help restore the symmetry at
high temperatures. As an example, we show in Fig-\ref{Vt} the corrected
thermal effective potential at such a temperature at which the unusual
behavior described in \cite{ELR} is supposed to appear, for example $T=1.7f$,
taking into account few order of the masses thermal corrections ($n=0,1,2$) in
(\ref{mto}).

As it is clear, it is enough to consider only the order ($n=2$) in
(\ref{mto}) to see that the EW symmetry is restored at this
temperature, $T=1.7f$. Here in this example, we have have chosen the
cut-off scale to be $\Lambda=1.3f$. If it is taken to be the NDA
value $\Lambda=4\pi f$, then one needs more corrections ($n>2$) to
show that the symmetry is restored, and for the order $n=2$, the EW
symmetry is restored at $T\simeq2.854f.$ This mean that the thermal
effective potential behavior shown in \cite{ELR}, is not an
intrinsic feature of LH models, but it appears due the incomplete
theory above such a validity scale. This validity scale of the
Littlest Higgs model could be estimated from the temperature range,
which in below the higher order corrections do not play a
significantly role in the electroweak dynamics. According to the
values of the $c_{n}$ parameters at the second temperature regime
$T>f$, where all particles are in thermal equilibrium, the series
(\ref{mto}) could be divergent! \cite{next}. But in the first
temperature regime, the $c_{n}$ parameters values are smaller than
the second regime case, and each term in the series (\ref{mto}) is
suppressed by an additional $T/f$ power. This naive estimation
allows us to suggest that the Little Higgs model is valid up to
$\Lambda\lesssim f$, and it is not useful to use the thermal
potential at temperatures above this scale. This could open a window
to investigate possible realization of a successful baryogenesis at
the weak scale within LH models.

In this letter, we have shown that the symmetry non-restoration effect at high
temperature for the Littlest Higgs model is due to periodicity of the
effective potential with respect to the field value and the largeness of the
Yukawa contribution to the thermal effective potential. This behavior
disappears when using novel corrections that are coming form the non-linear
nature the scalar representation. We have suggested also a cut-off for the
littlest Higgs $\Lambda\lesssim f$, which is consistent with the values
suggested from the unitarity study and the scalar loop analysis.

\vspace{0.5cm} \textit{I want to thank Christophe Grojean, Mikko
Laine and Salah Nasri for useful comments and suggestions. This work
was supported by the Algerian Ministry of Higher Education and
Scientific Research under the cnepru-project: D01720090023.}


\begin{thebibliography}{99}                                                                                               %
\bibitem {LHm}S. Weinberg, Phys. Rev. Lett. 29, 1698 (1972); H. Georgi and A.
Pais, Phys. Rev. D10, 539 (1974); Phys. Rev. D12, 508 (1975); D.B. Kaplan and
H. Georgi, Phys. Lett. B136, 183 (1984).

\bibitem {LH}N. Arkani-Hamed, A.G. Cohen, E. Katz and A.E. Nelson, J. High
Energy Phys. 07, 034 (2002).

\bibitem {phlh}T. Han, H.E. Logan, B. McElrath, and L.-T. Wang, Phys. Rev.
D67, 095004 (2003).

\bibitem {ELR}J.R. Espinosa, M. Losada and A. Riotto, Phys. Rev. D72, 043520 (2005).

\bibitem {CW}S. Coleman and E. Weinberg, Phys. Rev. D7, 1888 (1973).

\bibitem {esno}J.R. Espinosa and J.M. No, J. High Energy Phys. 01, 006 (2007).

\bibitem {Uni}S. Chang and H.-J. He, Phys. Lett. B586, 95 (2004).

\bibitem {Th}L. Dolan and R. Jackiw, Phys. Rev. D9, 3320 (1974); S. Weinberg,
Phys. Rev. D9, 3357 (1974).

\bibitem {TMa}B. Grinstein and M. Trott, J. High Energy Phys. 11, 064 (2008).

\bibitem {next}A. Ahriche, '\textit{The Littlest Higgs Electroweak Vacuum at
Finite Temperature}', in preparation.

\bibitem {ring}M.E. Carrington, Phys. Rev. D45 (1992) 2933.
\end{thebibliography}
\end{document}